\documentclass[dvips]{acta}
\usepackage{supertabular,lscape,epsfig}
\usepackage{amssymb}
\usepackage{amsmath}

\SetPages{0}{0}

\SetVol{61}{2011}

\begin{document}

\begin{Titlepage}
\Title{Refining Parameters of the XO-5 Planetary System with High-Precision Transit Photometry\footnote{Based on observations collected at the Centro Astron\'omico Hispano Alem\'an (CAHA), operated jointly by the Max-Planck Institut f\"ur Astronomie and the Instituto de Astrofisica de Andalucia (CSIC).}}
\Author{G.~~M~a~c~i~e~j~e~w~s~k~i$^{1,2}$, ~ M.~~S~e~e~l~i~g~e~r$^1$, ~ Ch.~~A~d~a~m$^1$, ~ St.~~R~a~e~t~z$^1$ ~ and ~ R.~~N~e~u~h~\"a~u~s~e~r$^1$}
{$^1$Astrophysikalisches Institut und Universit\"ats-Sternwarte, 
     Schillerg\"asschen 2--3, D--07745 Jena, Germany\\ 
     e-mail: gm@astri.uni.torun.pl\\
 $^2$Toru\'n Centre for Astronomy, N. Copernicus University, 
     Gagarina 11, PL--87100 Toru\'n, Poland
} 

\Received{January 7, 2011}
\end{Titlepage}

\Abstract{Studies of transiting extrasolar planets offer an unique opportunity to get to know the internal structure of those worlds. The transiting exoplanet XO-5~b was found to have an anomalously high Safronov number and surface gravity. Our aim was to refine parameters of this intriguing system and search for signs of transit timing variations. We gathered high-precision light curves of two transits of XO-5~b. Assuming three different limb darkening laws, we found the best-fitting model and redetermined parameters of the system, including planet-to-star radius ratio, impact parameter and central time of transits. Error estimates were derived by the prayer bead method and Monte Carlo simulations. Although system's parameters obtained by us were found to agree with previous studies within one sigma, the planet was found to be notable smaller with the radius of $1.03^{+0.06}_{-0.05}$ Jupiter radii. Our results confirm the high Safronov number and surface gravity of the planet. With two new mid-transit times, the ephemeris was refined to $\rm{BJD_{TDB}}$ $=$ $(2454485.66842 \pm 0.00028) + (4.1877537 \pm 0.000017) E$. No significant transit timing variation was detected.}{planetary systems -- stars: individual: XO-5 -- planets and satellites: individual: XO-5~b}

\section{Introduction}

Transiting exoplanets are of great importance for the astrophysics of extrasolar planetary systems. If the inclination of the planetary orbit is close to $90^{\circ}$, a planet periodically moves across a stellar disk, blocking a small fraction of the flux. This phenomenon -- a transit -- is observed as a small (up to $\sim$3\%) drop of host star's brightness. 
Combining spectroscopic and photometric data gives unique opportunity to determine planetary mass and radius, and hence the surface gravity and average density -- a key parameter for studying the internal structure of exoplanets. The first exoplanetary transits were observed for HD~209458~b (Charbonneau et al. 2000; Henry at al. 2000) which was discovered with the radial-velocity technique (Mazeh et al. 2000; Henry at al. 2000). The efficiency of photometric searches for transiting planets is seriously affected by the high fraction of false positive scenarios because variety of phenomena may mimic a planetary transit light curve. Examples of these false positives are central transit of a low-mass star in front of a large main-sequence star or red giant, grazing eclipses in systems comprising two main-sequence stars or a contamination of a fainter eclipsing binary along the same line of sight (Charbonneau et al. 2004). OGLE-TR-56~b is the first exoplanet discovered by the transit method. It was initially listed as a planetary candidate found by the Optical Gravitational Lensing Experiment (OGLE, Udalski et al. 1997) in the direction of the Galactic Centre (Udalski et al. 2002a,b). Then, the planetary nature of OGLE-TR-56~b was confirmed by radial velocity measurements by Konacki et al. (2003). Since then, over 100 extrasolar transiting planets have been discovered by numerous surveys. 

High-precision photometric follow-ups of transiting exoplanets allow to refine planetary and stellar parameters. Such observations may also lead to discoveries of additional, even very low mass bodies in extrasolar systems if transit time variations (TTVs) are detected (Miralda--Escud\'e 2002; Holman \& Murray 2005; Agol et al. 2005). The example of the Kepler 9 planetary system, in which there are at least 2 transiting planets, clearly demonstrates the usefulness of the TTV method (Holman et al. 2010). Preliminary detections of the TTV signal were reported for the WASP-3~b and WASP-10~b transiting planets (Maciejewski at al. 2010a,b). The source of these deviations from a strictly Keplerian case could be additional planets in these planetary systems. Detection of a TTV signal together with transit duration variations (TDVs) shifted in a phase by $\pi/2$ would indicate the presence of an exomoon of the transiting planet (Kipping 2009).

The star XO-5 (GSC 02959-00729, $\alpha=07^{\rm{h}}46^{\rm{m}}52^{\rm{s}}$, $\delta=+39^{\circ}05'41''$) was found to harbour a transiting hot Jupiter, XO-5~b, which orbits its host star within 4.2 days and whose mass and radius were found to be $M_{\rm{b}}=1.15\pm0.08$ $M_{\rm{J}}$ and $R_{\rm{b}}=1.15\pm0.12$ $R_{\rm{J}}$, respectively (Burke et al. 2008). High-resolution spectral observations revealed that the host star is a dwarf of the G8 spectral type, has the effective temperature of $5510\pm44$ K and is located $270\pm25$ pc from the Sun. The age of the system was estimated to be $8.5\pm0.8$ Gyr. The planet was independently confirmed by P\'al et al. (2009) who refined system's parameters. The planetary mass and radius were found to be $1.059\pm0.028$ $M_{\rm{J}}$ and $1.109\pm0.050$ $R_{\rm{J}}$, respectively. The host star was found to be slightly cooler with the effective temperature of $5370\pm70$ K and more evolved with the age of $14.8\pm2.0$ Gyr. The orbital period was refined to $4.187757\pm0.000011$ d and no sign of variations in transit timing was found. Using theoretical models, P\'al et al. (2009) found that the planet's core could be smaller than 10 Earth masses if a planet older than 4.5 Gyr is considered.

In this paper we present results of high-precision photometric follow-up observations whose aim was to refine parameters of the XO-5 system.

\section{Observations and data reduction}

Two transits of XO-5~b were observed with the 2.2-m telescope at Calar Alto Observatory (Spain) during 2 runs on 2010 January 21 and November 10. The Calar Alto Faint Object Spectrograph (CAFOS) in imaging mode was used as a detector. It was equipped with the SITe CCD matrix ($2048 \times 2048$, 24$\mu$m pixel, $0.53$ arcsec per pixel). A subframe limiting the field of view to $5.3 \times 3.7$ arcmin was used to shorten the read-out time. During the second run, binning in 2$\times$2 mode was used to additionally shorten the read-out time. The subframe was chosen in such a way to observe simultaneously a nearby comparison star GSC 02959-01873 which has a colour index and brightness similar to the ones of XO-5. This criterion was expected to minimise photometric trends caused by the differential atmospheric extinction. The photometric monitoring was performed in the Johnson $R$-band filter in which the instrument set-up is most sensitive. The telescope was significantly defocused and hence stellar profiles exhibited a donut-like shape (Fig.~1). This method minimises random and flat-fielding errors (e.g. Southworth et al.~2009). The stellar flux was spread over a ring of $\sim$17 arcsec in diameter. A visual inspection of the Digitized Sky Survey (DSS) images revealed no faint neighbour stars up to 40 arcsec in the direct vicinity of XO-5 nor the comparison star down to the limiting magnitude which is at least $R\sim20.5$ mag. The stellar images were kept exactly at the same position in the CCD matrix during each run thanks to auto guiding. Precise timing was assured by synchronising the computer's clock to Coordinated Universal Time (UTC) by Network Time Protocol software, accurate to better than 0.1 s.   

\begin{figure}[htb]
\begin{center}
\includegraphics[width=0.7\textwidth]{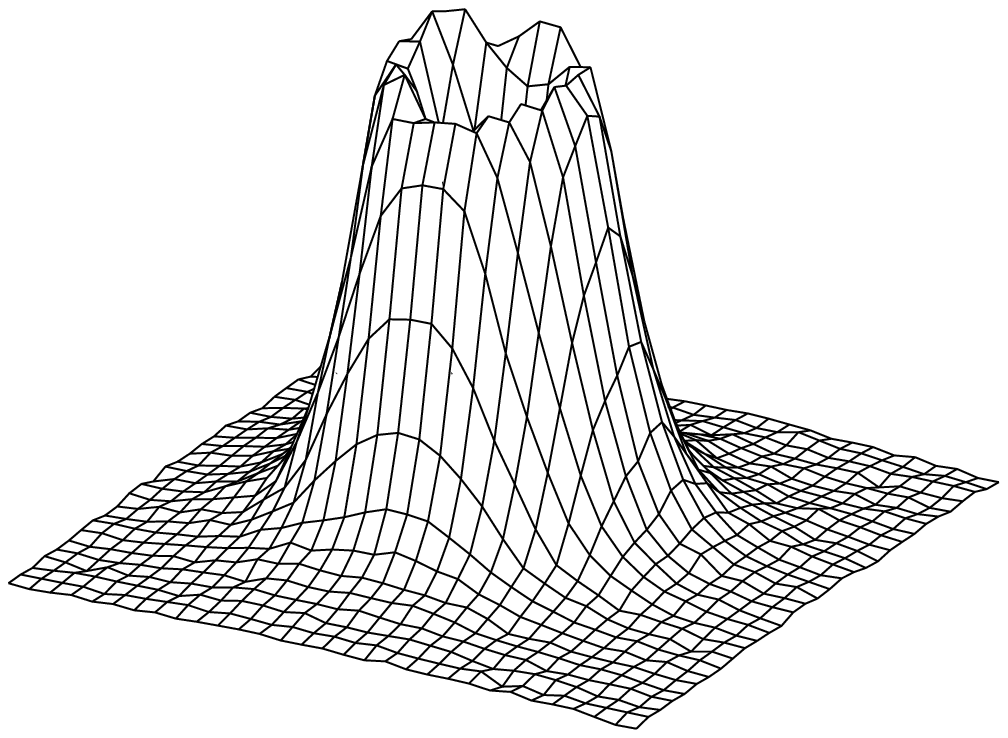}
\end{center}
\FigCap{The flux distribution of the stellar profile. The donut-like shape is caused by defocusing the telescope. The $30 \times 30$ pixel fragment corresponds to $15.9 \times 15.9$ arcsec on the sky (no binning). The logarithmic vertical scale was used for better contrast. The data were extracted from a typical 50-s exposure collected during the run on 2010 January 21.}
\end{figure}

CCD frames were processed using a standard procedure including debiassing but not flat-fielding which does not improve measurements in the case of defocused and auto-guided observations but may also degrade the quality of data (see e.g. Southworth et al. 2010). Indeed, our tests showed that dividing by a flat-field frame had no detectable effect on a final light curve. The magnitudes of XO-5 and comparison star were determined with differential aperture photometry. The aperture radii ranged from 8 to 18 pixels and 6 to 14 pixels for runs 1 and 2, respectively. Beginning from the aperture radii of 8 and 12 pixels for runs 1 and 2, respectively, the photometric scatter exhibited a flat minimum (or rather a plateau), in practice insensitive to greater aperture sizes. To avoid possible neighbour-star contamination, the light curve with minimal apertures for which the lowest scatter was achieved (i.e. 8 and 12 pixels for runs 1 and 2, respectively), were taken as the final ones. The comparison star was used to set the zero level of the photometric scale. No trends which could be approximated by a first- or second-order polynomial were detected in both transit light curves due to small changes of the airmass. Therefore, no detrending procedure was applied.

During the first run, sky conditions were non-photometric and data were acquired in gaps between clouds. The ingress phase was lost and egress data were significantly affected by thin clouds. Therefore, these data were found to be of marginal use. In the second run, almost photometric conditions allowed to record a complete transit light curve. Occasionally thin clouds affected flat-bottom phase of the transit. The exposure time was set to get roughly 1-min cadence and avoid saturation. It was refined during out-of-transit phase to achieve maximal efficiency. To avoid effecting transit timing, no exposure-time changes were done during ingress nor egress phases. Observations in both runs were carried out in dark time, just a few days before or after a new moon. Hence they were not affected by moonlight. The details of both observing runs are presented in Table~1.

\MakeTable{c c c c c c}{12.5cm}{The summary of observing runs: $N_{\rm{exp}}$ -- the number of useful exposures, $X$ -- airmass changes during a given run, $T_{\rm{exp}}$ -- exposure times. Dates are given in UT at the beginning of nights.}
{\hline
Run & Date & $N_{\rm{exp}}$ & $X$ & $T_{\rm{exp}}$ (s) & Binning mode\\ 
\hline
   1 & 2010 January 21 & 143 & $1.07\rightarrow1.00\rightarrow1.18$ & 45, 50, 60 & 1$\times$1  \\ 
   2 & 2010 November 10 & 190 & $1.16\rightarrow1.00\rightarrow1.02$ & 40, 45, 50 & 2$\times$2  \\
\hline
}

\section{Results}

Individual light curves were modelled with the JKTEBOP code (Southworth et al. 2004a, 2004b) which is based on the EBOP programme (Eclipsing Binary Orbit Program; Etzel 1981; Popper \& Etzel 1981). The software models both components of a system -- a planet and a host star -- as biaxial ellipsoids and performs a numerical integration in concentric annuli over the surface of each body to obtain the flux coming from the system. This feature of the code allows to avoid small and spherical planet approximations which are used in analytic light-curve generators based on Mandel \& Agol (2002), and hence to derive planet's oblateness. A model is fitted to the data by the Levenberg-Marquardt least-square procedure. The code converges rapidly toward a reliable solution and diminishes the correlation between fitted parameters (Southworth 2008). 

For the high-quality light curve from the second run, five parameters describing a shape of a light curve were allowed to float during fitting procedure. We used fractional radii of the host star and planet, defined as $r_{*}=\frac{R_{*}}{a}$ and $r_{\rm{b}}=\frac{R_{\rm{b}}}{a}$, respectively, where $R_{*}$ and $R_{\rm{b}}$ are the absolute radii of the bodies and $a$ is the orbital semi-major axis. In practise, the combinations of these parameters were used: a sum $r_{*}+r_{\rm{b}}$ and ratio $k=r_{*}/r_{\rm{b}}$ because they were found to be the less correlated with each other (Southworth 2008). The directly fitted orbital inclination, $i$, allowed to calculate the transit parameter $b=\frac{a}{R_{*}}\cos{i}$. The initial values of parameters listed above were taken from P\'al et al. (2009). To detect any variation in transit timing between observed transits, the mid-transit time was set as a free parameter whose initial value was calculated according to the ephemeris given by Burke et al. (2008). High-quality light curves need to set limb-darkening coefficients (LDCs) as free parameters and test different limb-darkening (LD) laws (Southworth 2008). Hence, we considered linear, logarithmic and square-root LD, for which theoretical LDCs in the Johnson $R$ band were bilinearly interpolated from tables by Van Hamme (1993). In a first iteration we allowed both linear $u$ and non-linear $v$ coefficients to vary but unrealistic values were derived. Therefore, we kept $v$ fixed at theoretical values and allow only $u$ to vary in final fitting runs. The contribution of $v$ into the error budget was included by perturbing it by $\pm0.1$ around its theoretical value and assuming a flat distribution.

The errors of derived parameters were determined in two ways for each combination of the data set and adopted LD law. Firstly, we run 1000 Monte Carlo (MC) simulations and a spread range of a given parameter within 68.3\% was taken as its error estimate. Secondly, the prayer-bead method (e.g. D\'esert et al. 2009; Winn et al. 2009) was used to check whether red noise is present in our data. MC errors were found to be 2--3 times smaller than the values returned by the prayer bead method, thus the latter ones were used as the final values. This finding indicates that beside Poisson noise there is correlated (red) noise in the light curve. 

\begin{figure}[htb]
\includegraphics[width=1.0\textwidth]{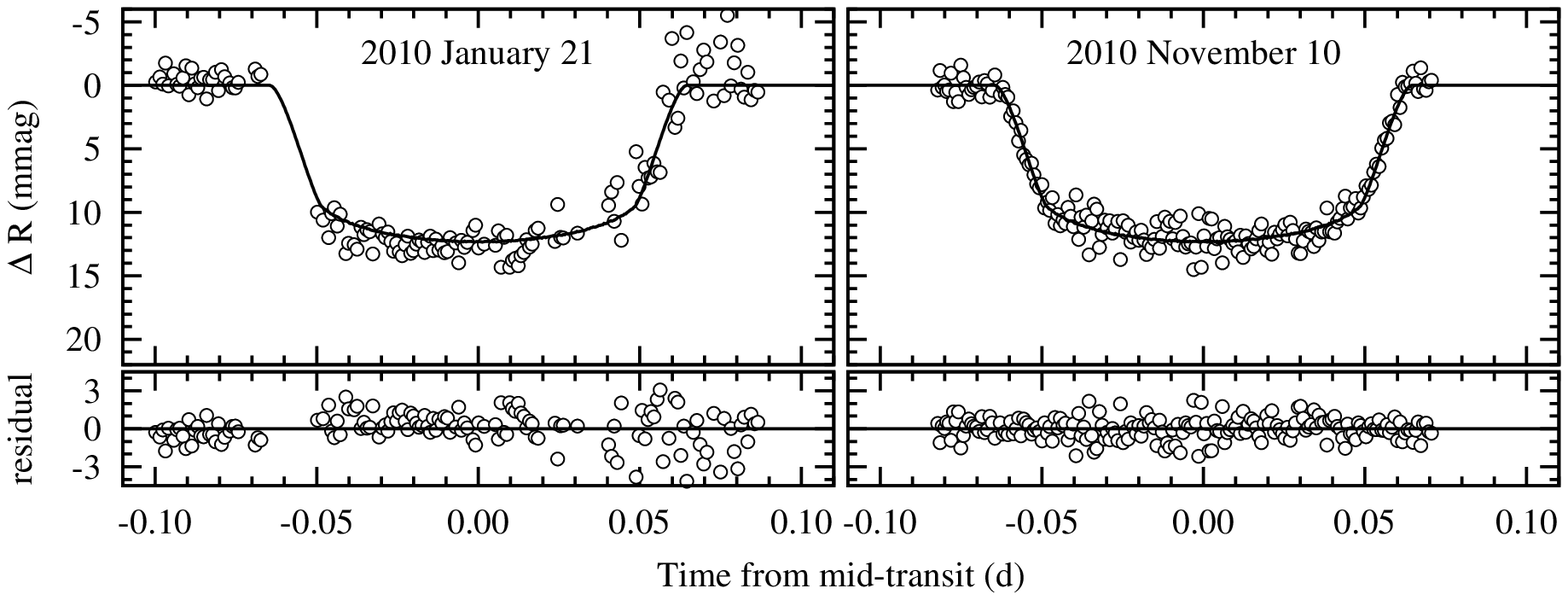}
\FigCap{Light curves of two transits of XO-5~b observed on 2010 January 21 and November 10. The best-fitting models, based on the square-root limb darkening law, are plotted with continuous lines. The residuals are shown in bottom panels.}
\end{figure}

\MakeTable{l c c c}{12.5cm}{Parameters of transit light-curve modelling derived for the high-quality light curve acquired in the second run on 2010 November 10. Linear, logarithmic and square-root limb darkening laws were considered. $T_0$ is based on UTC and is given as 2455511$+$. See text for details.}
{\hline
Parameter & Linear & Logarithmic & Square-root\\ 
\hline
$r_{*}+r_{\rm{b}}$      & $0.1112^{+0.0040}_{-0.0036}$    & $0.1136^{+0.0043}_{-0.0042}$    & $0.1102^{+0.0039}_{-0.0034}$    \\ 
$k$                     & $0.1018^{+0.0014}_{-0.0014}$    & $0.1018^{+0.0014}_{-0.0013}$    & $0.1013^{+0.0014}_{-0.0016}$    \\ 
$r_{*}$                 & $0.1009^{+0.0035}_{-0.0032}$    & $0.1032^{+0.0037}_{-0.0038}$    & $0.1000^{+0.0034}_{-0.0040}$    \\ 
$r_b$                   & $0.01028^{+0.00052}_{-0.00045}$ & $0.01050^{+0.00053}_{-0.00052}$ & $0.01013^{+0.00050}_{-0.00042}$ \\ 
$i$ (deg)               & $86.80^{+0.35}_{-0.38}$         & $86.6^{+0.4}_{-0.4}$            & $86.9^{+0.4}_{-0.4}$            \\ 
$b$ ($R_{*}$)           & $0.53^{+0.08}_{-0.08}$          & $0.55^{+0.09}_{-0.09}$          & $0.51^{+0.08}_{-0.08}$          \\ 
$u$                     & $0.45^{+0.07}_{-0.06}$          & $0.57^{+0.08}_{-0.09}$          & $0.12^{+0.07}_{-0.08}$          \\ 
$v$                     & $-$                             & $0.2171$$^a$                    & $0.5419$$^a$                    \\ 
$T_0$ $(\rm{JD_{UTC}})$ & $0.66486^{+0.00023}_{-0.00030}$ & $0.66487^{+0.00025}_{-0.00033}$ & $0.66483^{+0.00023}_{-0.00029}$ \\ 
$rms$ (mmag)            & $0.8334$                        & $0.8336$                        & $0.8329$                        \\ 
$\chi^2_{\rm{red}}$     & $1.6325$                        & $1.6326$                        & $1.6304$                        \\ 
\hline
\multicolumn{4}{l}{$^a$ permuted by $\pm0.1$ on a flat distribution}
}

Table~2 contains results obtained for the second-run light curve and individual LD laws. The sub-millimagnitude precision was achieved with the $rms$ of 0.83 mmag. The smallest $\chi^2_{\rm{red}}$ was achieved for the square-root LD law and this set of system's parameters was taken as the final one. However, it is worth noting that values of $rms$ of individual fits are comparable to each other what suggests that the choice of the LD law is not significant for the analysed light curve. Moreover, the timing errors seem to be independent on the choice of the LD law. In all three cases the precision between $20$ and $29$ s was achieved.

Full analysis of the first-run light curve was found to be unreliable, thus all parameters but mid-transit time were fixed and this light curve was used only for transit timing. The values of system's parameters were adopted from best-fitting model obtained for the second-run light curve. The error of mid-transit time was determined with the prayer bead method. The $\chi^2_{\rm{red}}$ of the fit was found to be 3.8 with the $rms=1.4$ mmag what reflects the lower quality of the data.   

The light curves acquired in both runs are plotted in Fig.~2 together with the best-fitting models and residuals\footnote{The data in a machine-readable form are available at http://web.astri.umk.pl/ttv.}.

The mid-transit times available in the literature and 2 new determinations reported in this paper were used to refine the transit ephemeris. We used 8 high-quality mid-transit times from Burke et al. (2008) rejecting unreliable ones which are based on the original XO photometry. We also used 4 mid-transit times given by P\'al et al. (2009); values for the individually fitted transit centres were taken. The mid-transit times were transformed from JD based on UTC into BJD based on Barycentric Dynamical Time (TDB) using the on-line converter\footnote{http://astroutils.astronomy.ohio-state.edu/time/utc2bjd.html} by Eastman et al. (2010). As a result of fitting a linear function of the epoch and period $P_{\rm{b}}$, we obtained:
\[
	T_{0}  = 2454485.66842 \pm 0.00028 \,  \; \rm{BJD_{TDB}} 
\]
\[
	P_{\rm{b}}  = 4.1877537 \pm 0.0000017 \,  \; \rm{d}. 
\]
Individual mid-transit errors were taken as weights. Results for new mid-transit times are summarised in Table~3. The observation minus calculation ($O-C$) diagram, plotted in Fig.~3, shows no significant deviation of data points from the new linear ephemeris. Some literature measurements deviate by more than 1 sigma which may be a result of underestimated timing errors.

\begin{figure}[htb]
\begin{center}
\includegraphics[width=1.0\textwidth]{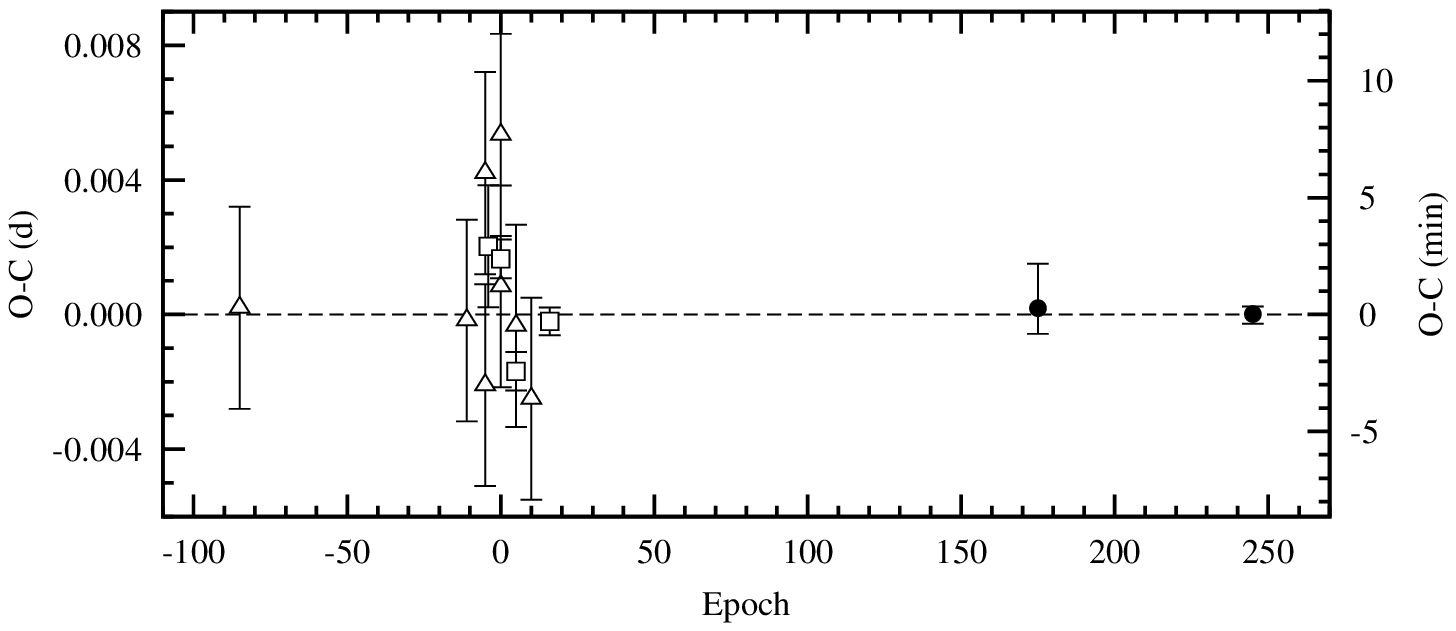}
\end{center}
\FigCap{The observation minus calculation ($O-C$) diagram for transit timing of XO-5~b, generated according to the new linear ephemeris. The open triangles and squares denote individual mid-transit times published by Burke et al. (2008) and P\'al et al. (2009), respectively. The filled symbols mark two mid-transit times reported in this paper. No transit timing variation was detected.}
\end{figure}

\MakeTable{c c c c c c}{12.5cm}{Results of transit timing. $T_{0}$ denotes the mid-transit times given as JD (based on Coordinated Universal Time, UTC) and BJD (based on Barycentric Dynamical Time, TDB). Errors of mid-transit times are in days. $O-C$ values were calculated according to the new ephemeris.}
{\hline
Run  & $T_{0}$ $(\rm{JD_{\rm{UTC}}})$ & $T_{0}$ $(\rm{BJD_{\rm{TDB}}})$ & $T_{0}$ error & Epoch & $O-C$ (d) \\ 
\hline
  1 & $2455218.5194$  & $2455218.5255$  & $^{+0.0013}_{-0.0008}$  & 175 & $+0.0002$  \\ 
  2 & $2455511.66483$ & $2455511.66809$ & $^{+0.00023}_{-0.00029}$ & 245 & $+0.00001$  \\
\hline
}

\section{Physical properties of XO-5 system}

Results of light-curve modelling allowed us to calculate planetary, stellar and geometrical parameters. We used results from the best-fitting model based on the second-run data and square-root LD law. $R_{\rm{b}}$ was calculated assuming $a=0.0488\pm0.0006$ AU (P\'al et al. 2009). The planetary mean density $\rho_{\rm{b}}$ uses XO-5~b's mass of $1.059\pm0.028$ $M_{\rm{J}}$, known from radial velocity measurements (P\'al et al. 2009). Surface gravitational acceleration, $g_{\rm{b}}$, was calculated according to a formula (Southworth et al. 2007): 
\begin{equation}
     g_{\rm{b}} = \frac{2\pi}{P_{\rm{b}}}\frac{\sqrt{1-e^2}}{r^{2}_{\rm{b}} \sin i} K_{*}\, , \; 
\end{equation}
where $P_{\rm{b}}$ is the orbital period of the planet (see Sect.~3) and $K_{*}$ is the stellar velocity amplitude taken from P\'al et al. (2009). The orbital eccentricity was assumed to be zero (Burke et al. 2008; P\'al et al. 2009). 

The equilibrium temperature, $T_{\rm{eq}}$, may be derived assuming the effective temperature of the host star derived by P\'al et al. (2009) and using the relation (Southworth 2010): 
\begin{equation}
  T_{\rm{eq}} = T_{\rm{eff}} \left(\frac{1-A}{4F}\right)^{0.25} \left( \frac{r_{*}}{2} \right)^{0.5}\, , \; 
\end{equation}
where $A$ and $F$ are the Bond albedo and heat redistribution factor, respectively. This formula was simplified by assuming relation $A=1-4F$ and the modified equilibrium temperature $T'_{\rm{eq}}$ was calculated (Southworth 2010). 

We calculated the Safronov number $\Theta$ which determines the efficiency with which a planet gravitationally scatters other bodies (Safronov 1972). This parameter is proportional to the ratio of the escape velocity from the planet and velocity of its orbital motion and can be calculated according to a formula (Southworth 2010):
\begin{equation}
  \Theta = \frac{M_{\rm{b}}}{M_{*}} r_{\rm{b}}^{-1}\, , \; 
\end{equation}
where the the value of $0.88\pm0.03$ $M_{\odot}$ was taken as star's mass $M_{*}$ (P\'al et al. 2009). 

Assuming that the planet's rotation and orbital periods are synchronised, the oblateness of the planet, defined as:
\begin{equation}
  f = 1 - \frac{r_1}{r_2}\, , \; 
\end{equation}
where $r_1$ and $r_2$ are the polar and equatorial radii, respectively, was found to be 0.0014. Such a small value indicates that any deviations of the XO-5~b's shape from an ideal sphere may be neglected.  

Results of calculations are collected in Table~4 in which literature determinations are also given for comparison. 

\MakeTable{l c c c}{12.5cm}{Physical properties of the XO-5 system derived from light-curve modelling. Listed parameters are explained in the text. Values derived by Burke et al. (2008) and P\'al et al. (2009) are given for comparison.}
{\hline
Parameter & This work & Burke et al. (2008) & P\'al et al. (2009)\\ 
\hline
\multicolumn{4}{c}{Planetary properties} \\
$R_{\rm{b}}$ $(R_{\rm{J}})$         & $1.03^{+0.06}_{-0.05}$       & $1.15\pm0.12$   & $1.109\pm0.050$           \\ 
$\rho_{\rm{b}}$ (g~cm$^{-3})$       & $1.27^{+0.19}_{-0.17}$       & $1.02\pm0.3$    & $0.96^{+0.14}_{-0.11}$    \\ 
$g_{\rm{b}}$ (m s$^{-2}$)           & $24.6^{+2.8}_{-2.3}$         & $22\pm5$        & $21.4^{+2.1}_{-1.9}$      \\ 
$T'_{\rm{eq}}$ (K)                  & $1201^{+36}_{-33}$           & $1244\pm48$     & $1221\pm27$               \\ 
$\Theta$                            & $0.114^{+0.013}_{-0.011}$    & $0.10\pm0.01$   & $0.105\pm0.005$           \\ 
$f$                                 & $0.0014$                     & $-$             & $-$                       \\ 
\multicolumn{4}{c}{Geometrical parameters} \\
$i$ (deg)                           & $86.9\pm0.4$                 & $86.8\pm0.9$    & $86.7\pm0.4$              \\
$b$ $(R_{*})$                       & $0.51\pm0.08$                & $0.55\pm0.09$   & $0.562^{+0.033}_{-0.052}$ \\ 
$R_{\rm{b}}/R_{*}$                  & $0.1013^{+0.0014}_{-0.0016}$ & $0.106\pm0.003$ & $0.1050\pm0.0009$         \\ 
\multicolumn{4}{c}{Stellar properties} \\
$R_{*}$ $(R_{\odot})$               & $1.05^{+0.05}_{-0.04}$       & $1.11\pm0.09$   & $1.08\pm0.04$             \\ 
$\rho_{*}$ $(\rho_{\odot})$         & $0.76^{+0.07}_{-0.06}$       & $0.72\pm0.14$   & $-$                       \\ 
$\log g_{*}$ (cgs)                  & $4.34^{+0.06}_{-0.05}$       & $4.34\pm0.07^a$ & $4.31\pm0.03$             \\ 
\hline
\multicolumn{4}{l}{$^a$ Here spectroscopic determination was taken.} \\
}

\section{Concluding discussion}
High-precision photometric follow-up observations of transiting planets are crucial for verifying physical properties of their planetary systems. Although system's parameters obtained by us were found to agree with previous studies by Burke et al. (2008) and P\'al et al. (2009) within one sigma, the planet was found to be notable smaller, i.e. by 7\%, comparing to the value obtained by P\'al et al. (2009). A shallower transit, and hence smaller ratio of planet and star radii, may be a sign of the presence of a third light contributing to the system's light curve, e.g. a faint star blended with XO-5 in the DSS image. To check this scenario, we explored the Naval Observatory Merged Astrometric Dataset (NOMAD, Zacharias et al. 2004) and found a faint source ($R=19.62$ mag), located 9~arcsec to the south from XO-5. Assuming an extreme and unlikely case, in which a total flux of the neighbour object contributed to the XO-5 light curve, the planetary radius would be affected by only 0.05\% -- over 100 times less than the detected difference. This result strengthens the reliability of our finding.   

The smaller planet's radius induces larger values of the mean density, $\rho_{\rm{b}}=1.27^{+0.19}_{-0.17}$ g~cm$^{-3}$, and gravitational acceleration, $g_{\rm{b}}=24.6^{+2.8}_{-2.3}$ m~s$^{-2}$, whose values are comparable to Jupiter's ones. Our results confirm the high Safronov number of XO-5~b which is equal to $0.114^{+0.013}_{-0.011}$. Such a high value suggests that the planet has not undergone a phase of the significant mass evaporation driven by the radiation of the host star (e.g. Hansen \& Barman 2007). This finding is not surprising if one considers the spectral type of the star (G8) and a relatively small amount of the extreme ultraviolet radiation that it emits. 

The transit times were found to be periodic within a 3-sigma level. This finding reveals no hint for the existence of the additional planet which could be discovered by the TTV method in the XO-5 system.

\Acknow{The authors are grateful to the staff of the Calar Alto Astronomical Observatory for their support during observing runs. GM acknowledges the financial support from the Polish Ministry of Science and Higher Education through the Iuventus Plus grant IP2010 023070. GM and SR acknowledge support from the EU in the FP6 MC ToK project MTKD-CT-2006-042514. CA would like to thank the German national science foundation Deutsche Forschungsgemeinschaft (DFG) for financial support in the programme NE 515/35-1. SR would like to thank the DFG for financial support in programmes NE 515/32-1 and NE 515/33-1. RN would like to thank the DFG for support in grants NE 515/32-1, 33-1, and 35-1. RN, SR, and CA acknowledge support by the DFG through SPP 1385: \textit{The first ten million years of the solar system -- a planetary materials approach}. GM, SR, and RN also acknowledge support from the DAAD PPP--MNiSW project 50724260--2010/2011 \textit{Eclipsing binaries in young clusters and planet transit time variations}. Finally, we would like to thank the DFG for financial support for the observing run on 2010 November 10 at Calar Alto in the programme NE 515/41-1. The authors are grateful to the anonymous referee for remarks improving the manuscript.}

\end{document}